\documentclass[a4paper, 11pt]{article}

\usepackage[pdftex, a4paper]{geometry}
\usepackage{graphicx}				
				
\usepackage{fancyhdr}
\usepackage{amsmath}
\usepackage{amsfonts}
\usepackage{dsfont}
\usepackage{mathrsfs}
\usepackage{amssymb}
\usepackage{bbm}
\usepackage{epsfig}
\usepackage[english]{babel}
\usepackage[all]{xy}

\newtheorem{thm}{Theorem}
\newtheorem{defn}{Definition}

\newtheorem{lem}{Lemma}
\newtheorem{remark}{Remark}

\title{\bf  Emergent Behaviors over Signed Random
Dynamical Networks: Relative-State-Flipping Model}
\date{}
\author{Guodong Shi, Alexandre Proutiere, Mikael  Johansson,\\ John. S. Baras, and Karl H. Johansson\thanks{G. Shi is with the College of Engineering and Computer Science, The Australian National University, Canberra ACT 0200, Australia.  A. Proutiere, M.  Johansson, and K. H. Johansson are with ACCESS Linnaeus Centre, Royal Institute of Technology, Stockholm 10044, Sweden. J. S. Baras is with Electrical and Computer Engineering, University of Maryland, College Park, MD 20742, USA. This work has been supported in part by the Knut   and Alice Wallenberg Foundation, the Swedish Research Council,   KTH SRA TNG, and  by AFOSR MURI grant FA9550-10-1-0573. e-mail: guodong.shi@anu.edu.au, alepro@kth.se, mikaelj@kth.se, baras@umd.edu, kallej@kth.se. }}
\begin{document}
\maketitle

\begin{abstract}
We study asymptotic dynamical patterns that emerge among a set of nodes interacting  in a
dynamically evolving signed random network, where positive links carry out standard consensus and  negative links induce relative-state flipping. A sequence of deterministic signed graphs define potential node  interactions that take place independently.  Each node receives a positive recommendation consistent with the standard consensus algorithm from its positive neighbors, and a negative recommendation defined by relative-state flipping  from its negative neighbors.  After receiving these recommendations, each node puts a deterministic weight to each recommendation, and then encodes these weighted recommendations in its state update through stochastic attentions defined by two Bernoulli random variables. We  establish a number of conditions regarding almost sure convergence and  divergence  of the node states. We also propose  a condition for  almost sure  state clustering for essentially weakly balanced graphs, with the help of several martingale convergence lemmas. Some fundamental differences on the impact of the deterministic weights and stochastic attentions to the node state evolution are highlighted between the current relative-state-flipping model and the state-flipping model considered in Altafini 2013 and Shi et al. 2014.
\end{abstract}


\section{Introduction}
The emergent behaviors, such as consensus, swarming, clustering, and learning,  of the dynamics evolving over a large complex network of interconnected nodes have attracted a significant amount of  research attention in the past decades \cite{degroot,vic95,jad03,swarm,naive}. In most cases node interactions are  collaborative, reflected by that
their state updates obey the same rule which is spontaneous or artificially  designed aiming for some particular collective task. This however  might not always be true since nodes  take on different, or even opposing, roles, where examples arise in biology  \cite{biologymath,nature13}, social science \cite{heider46,galam96,Antal2006}, and engineering \cite{barasgame}.

 Consensus problems  aim  to compute a weighted average of the initial values held by a collection of nodes, in a distributed manner.   The DeGroot's model \cite{degroot}, as a standard consensus algorithm, described  how opinions evolve in a network of agents, and showed that a simple deterministic opinion update based on the mutual trust and the differences in belief between interacting agents could lead to global convergence of the beliefs. Consensus dynamics have since then been widely adopted for describing opinion dynamics in social networks, e.g., \cite{naive,social2,misinformation}. In engineering sciences, a huge amount of literature has studied these algorithms for distributed averaging, formation  forming and load balancing between collaborative agents under fixed or time-varying interaction networks \cite{tsi,xiao,juliencdc,mor,ren,saber,caoming,julien13}. Randomized  consensus seeking has also been widely studied, motivated by  the random nature of interactions and updates in real complex networks~\cite{hatano,boyd,fagnani,kar2,touri,jad08,it10,TON13}.

This paper aims to study  consensus dynamics with both collaborative and non-collaborative node interactions.  A convenient framework for modeling different roles and relationships between agents is to use signed graphs  introduced in the classical work by Heider in 1946~\cite{heider46}. Each link is associated with a sign, either positive or negative, indicating collaborative or non-collaborative relationships. In \cite{altafini13}, a model for consensus over signed graphs was introduced for continuous-time dynamics, where a node  flips the sign of its true state to a negative (antagonistic)  node during the interaction. The author of \cite{altafini13} showed that state polarization (clustering) of the signed consensus model is closely related to the so-called structural balance in classical social signed graph theory~\cite{harary56}.  In \cite{shiJSAC}, the authors proposed a model for investigating the transition between agreement and disagreement when each link randomly takes three types of interactions: attraction, repulsion, and neglect, which was further generalized to a signed-graph setting in \cite{shiOR}.

We assume a sequence of deterministic signed graphs that defines the interactions of the network. Random node interactions take place under independent,  but not necessarily identically distributed,  random sampling of the environment.  Once interaction relations have been realized, each node receives a positive recommendation consistent with the standard consensus algorithm from its positive neighbors. Nodes receive negative recommendations from its negative neighbors.  After receiving these recommendations, each node puts a (deterministic) weight to each recommendation, and then encodes these weighted recommendations in its state update through stochastic attentions defined by two Bernoulli random variables.  In \cite{P1}, we studied almost sure convergence, divergence, and clustering under the definition of Altafini \cite{altafini13} for negative interactions, for which we referred to as a state-flipping model.

In this paper, we further investigate this random consensus model for signed networks under  a relative-state-flipping setting, where instead of taking  negative feedback  of the relative state in standard consensus algorithms \cite{degroot,jad03}, a positive feedback takes place along every  interaction arc of a negative sign. This relative-state flipping formulation is consistent with the models in \cite{shiJSAC,shiOR}, and can be viewed as a natural opposite of the DeGroot's type of node interactions. For the proposed relative-state-flipping model, we  establish a number of conditions regarding almost sure convergence and  divergence  of the node states. We also propose  a condition for  almost sure node state clustering for essentially weakly balanced graphs, with the help of several martingale convergence lemmas. Some fundamental differences on the impact of the deterministic weights and stochastic attentions to the node state evolution are highlighted between the current relative-state-flipping model and the state-flipping model.

The remainder of the paper is organized as follows. Section 2 presents the network dynamics and the node update rules, and specifies the information-level difference between the relative-state-flipping and state-flipping models.  Section 3 presents our main results; the detailed proofs are given in Section~4. Finally some concluding remarks are drawn in Section 5.

\vspace{3mm}
\subsection*{Graph Theory, Notations and Terminologies}

A simple directed graph (digraph) $\mathcal
{G}=(\mathcal {V}, \mathcal {E})$ consists of a finite set
$\mathcal{V}$ of nodes and an arc set
$\mathcal {E}\subseteq \mathcal{V}\times\mathcal{V}$, where  $e=(i,j)\in\mathcal {E}$ denotes   an
{\it arc}  from node $i\in \mathcal{V}$  to $j\in\mathcal{V}$ with $(i,i)\notin \mathcal{E}$ for all $i\in\mathcal{V}$.  We call node $j$  {\it reachable} from node $i$ if there is a directed path from $i$ to $j$. In particular every node is supposed to be reachable from itself. A node $v$ from which every  node in $\mathcal{V}$ is reachable is called a {\it center node} (root). A digraph $\mathcal{G}$ is {\it strongly connected} if every two  nodes are mutually reachable;  $\mathcal{G}$ has a spanning tree if it has a center node; $\mathcal{G}$ is {\it weakly connected} if  a connected undirected graph can be obtained by removing all the directions of the arcs in $\mathcal{E}$. A subgraph of $\mathcal
{G}=(\mathcal {V}, \mathcal {E})$, is a graph on the same node set $\mathcal {V}$ whose arc set is a subset of $\mathcal {E}$. The induced graph of $\mathcal{V}_i \subseteq \mathcal{V}$ on $\mathcal{G}$, denoted $\mathcal{G}|\mathcal{V}_i$, is the graph $(\mathcal{V}_i, \mathcal{E}_i)$ with $\mathcal{E}_i=(\mathcal{V}_i\times \mathcal{V}_i)\cap \mathcal{E}$. A weakly connected component of $\mathcal
{G}$ is a maximal weakly connected induced graph of $\mathcal
{G}$. If each arc $(i,j)\in\mathcal{E}$ is associated uniquely with a sign, either '$+$' or '$-$', $\mathcal {G}$ is called a signed graph and the sign of $(i,j)\in\mathcal {E}$ is denoted  as $\sigma_{ij}$.  The positive and negative subgraphs containing the positive and negative arcs of $\mathcal{G}$, are denoted as $\mathcal {G}^{+}=(\mathcal{V}, \mathcal{E}^+)$ and $\mathcal {G}^-=(\mathcal{V}, \mathcal{E}^-)$, respectively.

 Depending on the argument, $|\cdot|$ stands for the absolute value of a real number, the Euclidean norm of a vector or the cardinality of a set. The $\sigma$-algebra of a random variable is denoted as $\sigma(\cdot)$.   We use $\mathds{P}(\cdot)$ to denote the probability and $\mathds{E}\{\cdot\}$ the expectation
 of their arguments, respectively.


\section{Random Network Model  and Node Updates}

In this section, we present the considered random network model and specify individual  node dynamics. We use the same definition of random signed networks as introduced in \cite{P1}, where each link is associated with a sign indicating cooperative or antagonistic relations. In the current work we study relative-state-flipping dynamics along each negative arcs, in contrast with the state-flipping dynamics studied in \cite{P1}. The main difference of the information patterns between the two models will also be carefully explained.

\subsection{Signed Random Dynamical Networks}

Consider a network  with a set $\mathcal{V}=\{1,\dots,n\}$ of $n$ nodes, with $n\geq3$. Time is slotted for $t=0,1,\ldots$. Let $\big\{\mathcal {G}_t=(\mathcal{V},\mathcal{E}_t)\big\}_0^\infty$ be a sequence of (deterministic) signed directed graphs over node set $\mathcal{V}$. We denote by $\sigma_{ij}(t)\in\{+,-\}$ the sign of arc $(i,j)\in \mathcal{E}_t$. The positive and negative subgraphs containing the positive and negative arcs of $\mathcal{G}_t$, are denoted by $\mathcal {G}^{+}_t=(\mathcal{V}, \mathcal{E}^+_t)$ and $\mathcal {G}^-_t=(\mathcal{V}, \mathcal{E}^-_t)$, respectively. We say that the sequence of graphs $\{\mathcal {G}_t\}_{t\ge 0}$ is {\it sign consistent} if the sign of any arc $(i,j)$ does not evolve over time, i.e., if for any $s,t \ge 0$,
$$
(i,j)\in \mathcal {E}_s\ {\rm and}\ (i,j)\in \mathcal {E}_t\  \Longrightarrow \sigma_{ij}(s)=\sigma_{ij}(t).
$$
We also define $\mathcal {G}_\ast=(\mathcal{V},\mathcal{E}_\ast)$ with $\mathcal{E}_\ast=\bigcup_{t=0}^\infty\mathcal{E}_t$ as the total graph of the network. If  $\{\mathcal {G}_t\}_{t\ge 0}$ is sign consistent, then the sign of  each arc $\mathcal{E}_\ast$ never changes and in that case, $\mathcal {G}_\ast=(\mathcal{V},\mathcal{E}_\ast)$ is a well-defined signed graph. The notion of positive cluster in a signed directed graph is defined as follows.

\begin{defn} Let $\mathcal {G}$ be a signed digraph with positive subgraph $\mathcal {G}^+$.
 A subset $\mathcal {V}_\ast$ of the set of nodes $\mathcal {V}$ is a  positive cluster  if $\mathcal {V}_\ast$ constitutes a weakly connected component of $\mathcal {G}^{+}$. A positive cluster partition of $\mathcal {G}$ is a partition of $\mathcal{V}$ into $\mathcal{V}=\bigcup_{i=1}^{{\rm T}_{\rm p}} \mathcal{V}_i$ for some ${\rm T}_{\rm p}\geq 1$, where for all $i=1,\ldots, {\rm T}_{\rm p}$, $\mathcal{V}_i$ is a  positive cluster.
 \end{defn}
Note that  $\mathcal {G}$ admitting a positive-cluster partition is a generalization of the classical definition of weakly structural balanced graph for which negative links are strictly forbidden inside each positive cluster \cite{davis63}.  From the above definition, it is clear that for any signed graph $\mathcal {G}$, there is a unique positive cluster  partition $\mathcal{V}=\bigcup_{i=1}^{{\rm T}_{\rm p}} \mathcal{V}_i$ of $\mathcal{G}$, where ${\rm T}_{\rm p}$ is the  number of positive clusters covering the entire set ${\cal V}$ of nodes.

\begin{figure*}[t]
\begin{center}
\includegraphics[height=2.0in]{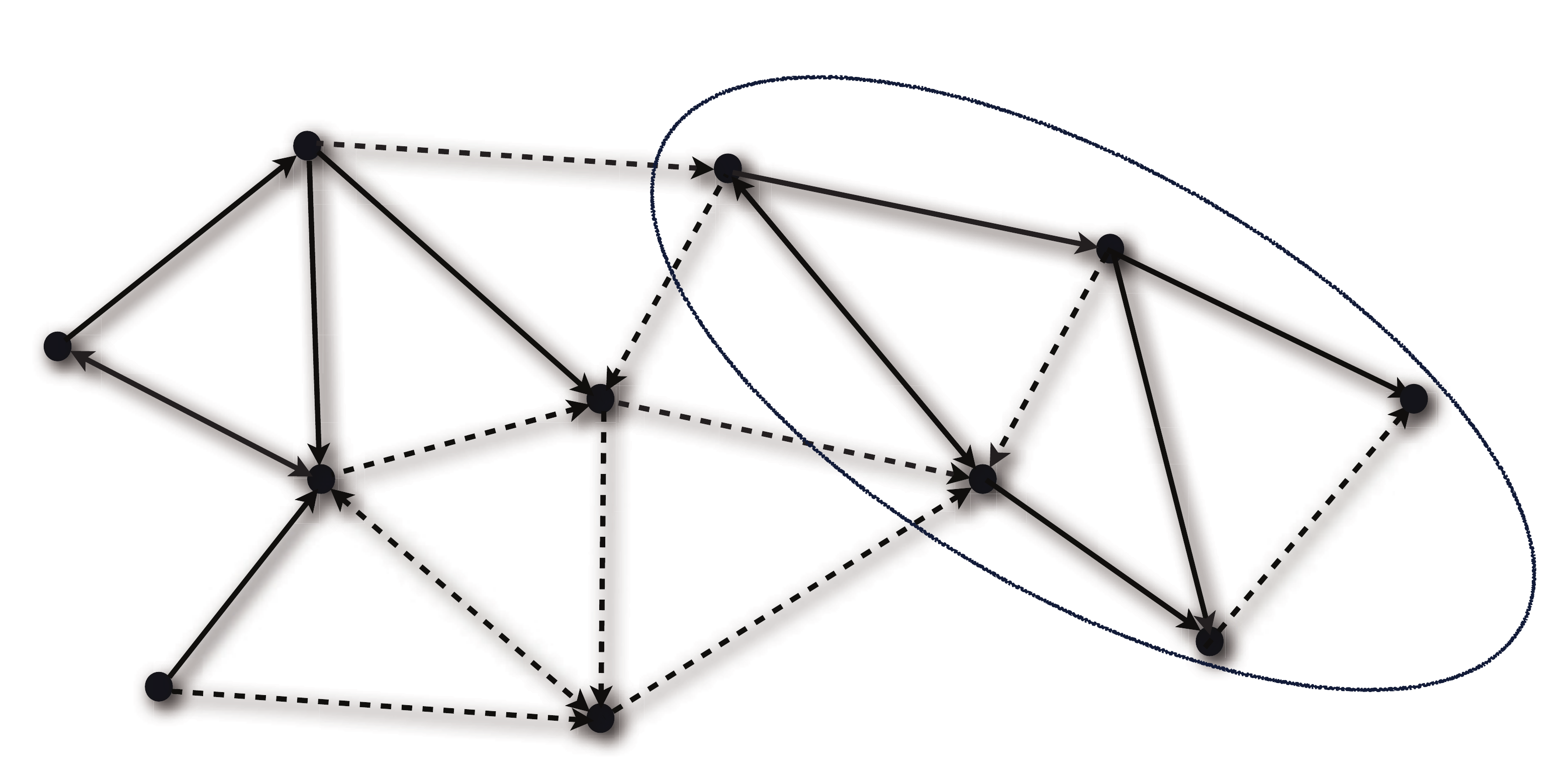}
\caption{A signed network and its three positive clusters. The  positive arcs are solid, and the negative arcs are dashed. Note that negative arcs are allowed within  positive clusters. } \label{frontier}
\end{center}
\end{figure*}

Each node randomly interacts with her neighboring nodes in ${\cal G}_t$ at time $t$. We present a general model on the random node interactions  at a given time $t$. At time $t$, some pairs of nodes are randomly selected for interaction. We denote by $E_t\subset {\cal E}_t$ the random subset of arcs corresponding to interacting node pairs at time $t$. To be precise, $E_t$ is sampled from the distribution $\mu_t$ defined over the set $\Omega_t$ of all subsets of arcs in ${\cal E}_t$. We assume that $E_0,E_1,\ldots$ form a sequence of independent sets of arcs. Formally, we introduce the probability space $(\Theta,\mathcal{F},\mathrm{P})$ obtained by taking the product of the probability spaces $(\Omega_t,{\cal S}_t,\mu_t)$, where $\mathcal{S}_t$ is the discrete $\sigma$-algebra on $\Omega_t$: $\Theta=\prod_{t\ge 0}\Omega_t$, ${\cal F}$ is the product of $\sigma$-algebras ${\cal S}_t$, $t\ge 0$, and $\mathrm{P}$ is the product probability measure of $\mu_t, t\ge 0$. We  denote by $G_t=(\mathcal{V}, E_t)$ the random subgraph of ${\cal G}_t$ corresponding to the random set $E_t$ of arcs. The disjoint  sets $E^+_t$ and $E^-_t$ denote the positive and negative arc set of $E_t$, respectively. Finally, we split the random set of nodes interacting with node $i$ at time $t$ depending on the sign of the corresponding arc: for node $i$, the set of positive neighbors is defined as ${N}^+_i(t):= \big\{j: (j,i)\in E_t^+\big\}$, whereas similarly, the set of negative neighbors is ${N}^-_i(t):= \big\{j: (j,i)\in E_t^-\big\}$.

\subsection{Node updates}
Each node $i$ holds a state $s_i(t)\in \mathds{R}$ at $t=0,1,\dots$. To update her state at time $t$, node $i$ considers recommendations received from her positive and negative neighbors:
\begin{itemize}
 \item[(i)] The positive  recommendation node $i$ receives  at time $t$ is
$$
h_i^+(t):=-\sum_{j\in{N}_i^+(t) } \big(s_i(t)-s_j(t)\big);
$$
\item[(ii)] The negative recommendations node  $i$ receives at time $t$ is defined as: $$
h_i^-(t):= \sum_{j\in{N}_i^-(t) } \big(s_i(t)-s_j(t)\big).
$$
\end{itemize}
In the above expressions, we use the convention  that summing over empty sets yields a recommendation equal to zero, e.g., when node $i$ has no positive neighbors, then $h_i^+(t)=0$. In view of the definition of $h_i^-(t)$ in contrast to $h_i^+(t)$, the model is referred to as the  relative-state-flipping model.

\begin{remark}
In \cite{P1},  we have considered another notion of negative recommendations, namely the state-flipping model introduced in \cite{altafini13},  defined as $h_i^-(t):= -\sum_{j\in{N}_i^-(t) } \big(s_i(t)+s_j(t)\big)$. We remark that for the relative-state-flipping model, the network does not require a central global coordinate system and nodes can interact based on relative state only. As has been pointed in \cite{P1}, in the state-flipping model, the network nodes are necessary to share a common knowledge  of the origin of the state space.
\end{remark}

\begin{remark}
The two definitions of negative recommendations, the relative-state-flipping model considered in the current paper, and the state-flipping model studied in \cite{altafini13, P1},  have  different physical interpretations and make different assumptions on the knowledge that nodes possess  about their neighbor relationships.  In the state-flipping model,  naturally  it is the {\it head} node along each negative arc, that possesses the knowledge of sign of that arc.  In the relative-state-flipping model, on the other hand, it is the {\it tail} node knows the sign of each directed arc so that nodes  know if a specific neighbor is positive or negative to implement the state updates that cause the  repulsive influence from its negative neighbors.
\end{remark}

Let $\{{B}_t\}_{t\ge 0}$ and $\{{D}_t\}_{t\ge 0}$ be two sequences of independent Bernoulli random variables. We assume that $\{{B}_t\}_{t\ge 0}$, $\{{D}_t\}_{t\ge 0}$, and $\{G_t\}_{t\ge 0}$ define independent processes. For any $t\ge 0$, define $b_t=\mathds{E}\{B_t\}$ and $d_t=\mathds{E}\{D_t\}$. The processes $\{{B}_t\}_{t\ge 0}$ and $\{{D}_t\}_{t\ge 0}$ represent how much  attention node $i$ pays to  the positive and negative recommendations, respectively. Node $i$ updates her state as follows:
\begin{align}\label{9}
 s_i(t+1)=s_i(t)+  \alpha{B}_t h_i^+(t) +\beta{D}_t h_i^-(t),
\end{align}
where $\alpha,\beta>0$ are two positive constants marking the  weight each node put on the positive and negative recommendations, respectively.

Let  $s(t)=\big(s_1(t)\dots s_n(t)\big)^T$ be the random vector representing the network state  at time $t$. The main objective of this paper is to analyze the behavior of the stochastic process $\{s(t)\}_{t\ge 0}$. In the following, we denote by $\mathds{P}$ the probability measure capturing all random components driving the evolution of $s(t)$.

In the remainder of the paper, we establish the asymptotic properties of the network state evolution under relative-state-flipping model. As will be shown in the following, the state-flipping and relative-state-flipping models share some common nature, e.g., almost sure state convergence/divergence, no-survivor property, etc. In the mean time these common properties can be driven by fundamentally different parameters regarding  network connectivity and recommendation weights and attentions. For the consistency of presentation we introduce the same set of assumptions on the random graph process and the connectivity of the dynamical environment as used in~\cite{P1}.

\vspace{2mm}

\noindent {\bf A1.} There is a constant  $p_\ast \in (0,1)$ such that for all $t\geq0$ and  $i,j\in \mathcal{V}$, $\mathds{P}\big( (i,j)\in E_t \big)\geq p_\ast$ if $(i,j)\in \mathcal{E}_t$.

\noindent {\bf A2.} There is an integer $K \geq 1$ such that the union graph $\mathcal{G}\big({[t,t+K-1]}\big)=\big(\mathcal{V}, \bigcup_{\tau \in [t,t+K-1] } \mathcal{E}_\tau\big)$ is strongly connected for all $t\geq0$.

\noindent
{\bf A3.}  $\{\mathcal{G}_t\}_{t\ge 0}$ is sign consistent admitting  a  total graph  $\mathcal{G}_\ast$.

\noindent {\bf A4.}  There is an integer $K\geq 1$ such that the union graph $\mathcal{G}^+\big({[t,t+K]}\big)=\big(\mathcal{V}, \bigcup_{\tau \in [t,t+K-1] } \mathcal{E}_\tau^+\big)$ is strongly connected for all $t\geq0$.

\noindent {\bf A5.}  There is an integer $K\geq 1$ such that the union graph $\mathcal{G}^-\big({[t,t+K]}\big)=\big(\mathcal{V}, \bigcup_{\tau \in [t,t+K-1] } \mathcal{E}_\tau^-\big)$ is strongly connected for all $t\geq0$.

\noindent {\bf A6.} The events  $\{ (i,j)\in G_t \}$, $i,j\in\mathcal{V}$, $t=0,1,\dots$ are independent and there is a constant  $p^\ast \in (0,1)$ such that for all $t\geq0$ and  $i,j\in \mathcal{V}$, $\mathds{P}\big( (i,j)\in G_t \big)\leq  p^\ast$ if $(i,j)\in \mathcal{E}_t$.

\vspace{2mm}

\section{Main Results}
In this section, we present the main results for the  asymptotic behaviors of the random process defined by the considered relative-state-flipping model.

\subsection{General Conditions}
First of all, the following theorem provides general conditions for convergence and divergence.

\medskip

\begin{thm}\label{thmrsr1} Let A1 hold and  $\alpha \in (0, (n-1)^{-1})$ and  $\beta>0$. Assume that for any $t\ge 0$, $\mathcal{G}_t\equiv \mathcal{G}   $ for some digraph $\mathcal{G}$, and that each positive cluster of $\mathcal{G}  $ admits a spanning tree in $\mathcal{G}  ^+$. For Algorithm (\ref{9}) under the relative-state-flipping model, we have:
\begin{itemize}
\item[(i)] If $\sum_{t=0}^\infty d_t <\infty$, then $\mathds{P} \big( \lim_{t\rightarrow \infty}s_i(t) \ {\rm exits}\big)=1$ for all node $i\in {\cal V}$ and all initial states $s(0)$;\\
\item[(ii)] If $\sum_{t=0}^\infty d_t =\infty$,  $\mathcal{G}  $ has  two positive clusters with no negative links in each cluster, and there is a negative arc between any two nodes from different clusters,   then there exist an infinite number of initial states $s(0)$ such that
\begin{align}\label{18}
\mathds{P} \big( \lim_{t\rightarrow \infty}\max_{i,j\in \mathcal{V}} | s_i(t)-s_j(t)|=\infty\big)=1.
\end{align}
\end{itemize}
\end{thm}

\medskip

The first part of the above theorem indicates that when the environment is frozen, and when positive clusters are properly  connected, then irrespective of the mean of the positive attentions $\{b_t\}_0^\infty$, the system states converge if the attention each node puts in her negative neighbors  decays sufficiently fast over time. The second part of the theorem states that when this attention does not decay, divergence can be   observed.

\begin{remark} Theorem \ref{thmrsr1} shows that well-structured positive arcs and asymptotically decaying attention guarantee state convergence for relative-state-flipping model. The essential reason is that when $\sum_{t=0}^\infty d_t <\infty$, the first Borel-Cantelli lemma (cf. Theorem 2.3.1, \cite{durr}) ensures that along almost every sample path, negative interactions happen only for a finite number of time instants. The positive interactions continue to guide the network states to a finite limit under suitable connectivity.  It is then clear that the same condition can also guarantee state-convergence for the state-flipping model considered in \cite{P1}.

In fact, for the state convergence property of the state-flipping model, a much stronger conclusion regarding state convergence  was shown  (Theorem 1 in \cite{P1}) indicating  that each positive/negative arc contributes to the state convergence under constant attention $\{b_t\} $ and $\{d_t\}$, as long as $\alpha+\beta \leq (n-1)^{-1}$. We can easily  build examples showing that it is a completely different story on this matter for the relative-state-flipping model considered in the current paper.
\end{remark}

 \begin{remark}The divergence statement in Theorem \ref{thmrsr1} is not true for  the state-flipping model \cite{P1},  where almost sure state divergence always requires sufficiently large $\beta$.
\end{remark}

\subsection{Deviation Consensus}
Next, we provide a sufficient condition for almost sure deviation consensus as defined below.

\medskip

\begin{defn}
Algorithm (\ref{9}) achieves almost sure deviation consensus if $$
\mathds{P} \big( \limsup_{t\rightarrow \infty} \max_{i,j\in \mathcal{V}} | s_i(t)-s_j(t)| = 0\big)=1.
$$
\end{defn}

\medskip

Note that almost sure deviation consensus means that the distances among the node states converge to zero, but convergence of each node state is not required.  We need the following assumption, which is a relaxed version of Assumption A4.

\medskip

\noindent {\bf A7.}   There is an integer $K\geq 1$ such that the union graph $\mathcal{G}^+\big([t,t+K]\big)=\big(\mathcal{V}_i, \bigcup_{\tau \in [t,t+K-1] } \mathcal{E}_\tau^+\big)$ has a spanning tree for all $t\geq0$.

\medskip

\begin{thm}\label{thmrsr2}  Assume that  A1 and A7 hold and that $\alpha \in (0,(n-1)^{-1})$. Denote $K_0=(2n-3)K$ and $\rho_\ast=\min\{\alpha, 1-(n-1)\alpha\}$. Define
$$
X_m=\frac{p_\ast^{n-1} \rho_\ast^{K_0}}{2}  \prod_{t=mK_0}^{(m+1)K_0-1}\big(b_t(1-d_t)\big),
 $$
 and $$
 Y_m=\big(1+2\beta(n-1) \big)^{K_0}\big(1- \prod_{t=mK_0}^{(m+1)K_0-1}(1-d_t)\big).
 $$
 Then under the relative-state-flipping model, if $0\leq  X_m-Y_m \leq 1$ for all $m\geq 0$ and  $\sum_{m=0}^\infty (X_m-Y_m)=\infty$, Algorithm (\ref{9}) achieves almost sure deviation consensus for  all initial states.
\end{thm}

\medskip

 \begin{remark}
A direct consequence of Theorem \ref{thmrsr2} is that  if  $b_t\equiv b  $ and $d_t\equiv d$ with $b  ,d \in (0,1)$ and $\beta>0$, there exists $d_\star >0$ such that whenever $d<d_\star$, deviation consensus is achieved almost surely. Observe that   deviation consensus  does not necessarily guarantee the convergence of the state of each node. In fact, simple examples can be constructed with arbitrarily small $\beta$ such that under the relative-state-flipping model, the state of each node grows arbitrarily large while deviation consensus still holds. This contrasts  the result for the state-flipping model:  the condition  $\alpha+\beta<(n-1)^{-1}$   prevents the state of individual nodes to diverge (Theorem 1 in \cite{P1}).
\end{remark}

\subsection{Almost Sure Divergence}
We continue to provide conditions under which the maximal gap between the states of two nodes grows large almost surely, and establish a no-survivor property. We introduce a new connectivity condition on the negative graph, which is a relaxed version of Assumption A5.

\medskip

\noindent {\bf A8.}  There is an integer $K\geq 1$ such that the union graph $\mathcal{G}^-\big({[t,t+K]}\big)=\big(\mathcal{V}, \bigcup_{\tau \in [t,t+K-1] } \mathcal{E}_\tau^-\big)$ is weakly connected for all $t\geq0$.

\medskip

\begin{thm}\label{thmrsr3} Assume that A1, A6, and A8 hold and that $\alpha \in [0,(n-1)^{-1}/2)$. Let $b_t\equiv b  $ and $d_t\equiv d$ for some constants $b  ,d \in (0,1)$. Let $\beta >0$ and fix $d$. Then for Algorithm (\ref{9}) under the relative-state-flipping model, there is $b_\star >0$ such that whenever $b  <b_\star$, we have    $
\mathds{P} \big( \lim_{t\rightarrow \infty} \max_{i,j\in \mathcal{V}} | s_i(t)-s_j(t)| = \infty\big)=1
$
for almost all initial states (under the standard Lebesgue measure).
\end{thm}

\medskip


\begin{remark}
Theorem \ref{thmrsr3} indicates that in relative-state-flipping model, almost sure relative-state divergence can be achieved as long as negative interactions happen sufficiently more often than the positive interactions. As explained in above remarks, for state-flipping model, state divergence necessarily require sufficiently large weight on negative recommendations.
\end{remark}

%
%

\subsection{State Clustering}
Finally, we investigate the clustering of states of nodes within each positive cluster.

\medskip

\noindent {\bf A9.} Assume that A3 holds and let $\mathcal{V}=\bigcup_{i=1}^{{{\rm T}_{\rm p}}} \mathcal{V}_i$ be a positive-cluster partition of the total graph $\mathcal{G}_\ast$. There is an integer $K\geq 1$ such that the union graph $\mathcal{G}^+\big({[t,t+K]}\big)\big|_{\mathcal{V}_i}=\big(\mathcal{V}_i, \bigcup_{\tau \in [t,t+K-1] } \mathcal{E}_\tau^+\big|_{\mathcal{V}_i}\big)$ has a spanning tree for all $t\geq0$.

\medskip

\begin{thm} \label{thmrsr4} Assume that A1, A3 and A9 hold and let $\mathcal{V}=\bigcup_{i=1}^{{{\rm T}_{\rm p}}} \mathcal{V}_i$ be a positive-cluster partition of $\mathcal{G}_\ast$. Let $\alpha \in (0,(n-1)^{-1})$. Define $J(m)=\prod_{t=mK_0}^{(m+1)K_0-1}b_t$ and $W(m)=\sum_{t=mK_0}^{(m+1)K_0-1} d_t$ with $K_0=(2n-3)K$. Further assume that $\sum_{m=0}^\infty J(m)=\infty$, $\sum_{t=0}^\infty d_t   <\infty$, and $\lim_{m \rightarrow\infty}W(m)/J(m)=0$. Then under the relative-state-flipping model, for any initial state $s(0)$, Algorithm (\ref{9}) achieves a.s. state clustering in the sense that there are ${{\rm T}_{\rm p}}$ real-valued random variables, $w^\ast_1,\dots,w^\ast_{{{\rm T}_{\rm p}}}$, such that $$
\mathds{P} \Big( \lim_{t\rightarrow \infty }s_i(t)= w^\ast_j,\ i\in\mathcal{V}_j,\ j=1,\dots, {{\rm T}_{\rm p}}\Big)=1.
$$
\end{thm}

\medskip

Theorem \ref{thmrsr4} shows the possibility of state clustering for  every positive cluster, whose proof is based on a martingale convergence lemma.
\section{Proofs of Statements}
In this section, we establish the proofs of the various statements presented in the previous section.
\subsection{Supporting Lemmas}

We list three martingale convergence lemmas (see e.g. \cite{polyak}), and a result that will be instrumental in the analysis of the system convergence under the relative-state-flipping model.

\medskip

\begin{lem}\label{lem6} Let $\{v_t\}_{t\ge 0}$ be a sequence of non-negative random variables with $\mathds{E}\{v_0\}<\infty$. Assume that for any $t\ge 0$,
$$
\mathds{E}\{v_{t+1}|v_0,\dots,v_t\}\leq (1+\xi_t)v_t+\theta_t,
$$
where $\{\xi_t\}_{t\ge 0}$ and $\{\theta_t\}_{t\ge 0}$ are two (deterministic) sequences of non-negative numbers satisfying $\sum_{t=0}^\infty {\xi_t}<\infty$ and $\sum_{t=0}^\infty {\theta_t}<\infty$. Then $\lim_{t\to\infty} v_t = v$ a.s. for some random variable $v\geq 0$.
\end{lem}

\medskip

\begin{lem}\label{lem7}
Let$\{v_t\}_{t\ge 0}$ be a sequence of non-negative random variables with $\mathds{E}\{v_0\}<\infty$. Assume that for any $t\ge 0$,
$$
\mathds{E}\{v_{t+1}|v_0,\dots,v_t\}\leq (1-\xi_t)v_t+\theta_t,
$$
where $\{\xi_t\}_{t\ge 0}$ and $\{\theta_t\}_{t\ge 0}$ are two (deterministic) sequences of non-negative numbers satisfying $\forall t\ge 0$, $0\leq \xi_t \leq 1$, $\sum_{t=0}^\infty {\xi_t}=\infty$, $\sum_{t=0}^\infty {\theta_t}<\infty$, and $\lim_{t\to\infty}{\theta_t}/{\xi_t}=0$. Then $\lim_{t\to\infty} v_t = 0$ a.s..
\end{lem}

\medskip

\begin{lem}\label{lem8}
Let $\{v_t\}_{t\ge 0},\{\xi_t\}_{t\ge 0},\{\theta_t\}_{t\ge 0}$ be sequences of non-negative random variables. Assume that for any $t\ge 0$,
$$
\mathds{E}\{v_{t+1}|\mathcal{F}_t\}\leq (1+\xi_t)v_t+\theta_t,
$$
where $\mathcal{F}_t=\sigma(v_0,\dots,v_t;\xi_0,\dots,\xi_t;\theta_0,\dots,\theta_t)$. Suppose $\sum_{t=0}^\infty {\xi_t}<\infty$ and $\sum_{t=0}^\infty {\theta_t}<\infty$ almost surely. Then $\lim_{t\to\infty} v_t = v$ a.s. for some random variable $v\geq 0$.
\end{lem}

\medskip

We define $h(t):= \min_{i\in\mathcal{V}} s_i(t)$, $H(t):= \max_{i\in\mathcal{V}} s_i(t)$, and $\mathcal{H}(t):=H(t)-h(t)$, which will be used throughout the rest of the paper. The following lemma holds.

\medskip

\begin{lem}\label{lem10}
 Assume that $\alpha \in [0, (n-1)^{-1}]$ and that $\sum_{t=0}^\infty d_t<\infty$. Then  under the relative-state-flipping model, for all initial states, each of $h(t)$, $H(t)$, $\mathcal{H}(t)$ converges almost surely.
\end{lem}

\noindent{\it Proof.} We build the proof in steps.

\medskip

\noindent Step 1. In this step, we prove the convergence of $\mathcal{H}(t)$. Since $\alpha \in [0, (n-1)^{-1}]$,  the proposed algorithm simply does weighted  averaging  when  $D_t=0$.  It is therefore well known that $H(t+1)\leq H(t)$, $h(t+1)\geq h(t)$, and  $\mathcal{H}(t+1)\leq \mathcal{H}(t)$ if $D_t=0$. On the other hand, when $D_t=1$, it holds from the structure of the algorithm that $\mathcal{H}(t+1)\leq (2\beta (n-1)+1)\mathcal{H}(t) $. We deduce that:
\begin{align}\label{100}
\mathds{E}\big\{\mathcal{H}(t+1)|\mathcal{H}(t)\big\}\leq \big(1+ 2\beta (n-1)d_t\big)\mathcal{H}(t),
\end{align}
which, in view of Lemma \ref{lem6}, implies that $\mathcal{H}(t)\rightarrow \mathcal{H}_\ast$ almost surely for some $\mathcal{H}_\ast \geq 0$.

\medskip

\noindent Step 2. Now for $H(t)$, we easily see from (\ref{100}) that
\begin{align}\label{r1}
\mathds{E}\big\{{H}(t+1)|{H}(t)\}\leq H(t)+\big(1+ 2\beta (n-1)d_t\big)\mathcal{H}(t).
\end{align}
Since we have proved that $\mathcal{H}(t)$ converges a.s. and $\sum_td_t<\infty$, we deduce that $\sum_t\big(1+ 2\beta (n-1)d_t\big)\mathcal{H}(t)<\infty$ a.s.. Further, in light of the first Borel-Cantelli Lemma (cf. Theorem 2.3.1, \cite{durr}), $\sum_td_t<\infty$ ensures that
$$
\mathds{P}\Big( \liminf_{t\rightarrow \infty} H(t)>-\infty\Big)=1
$$
because $H(t)\geq h(t)$ and $$
\big\{h(t+1)<h(t)\big \} \subseteq \big\{D_t=1\big \}
$$
for any $t\geq 0$. Thus, $\bar{H}(t):={H}(t)-\inf_{t\geq 0} H(t)$ is a well-defined nonnegative random variable for any $t\geq0$, and (\ref{r1}) implies
   \begin{align}\label{r2}
\mathds{E}\big\{\bar{{H}}(t+1)|\bar{{H}}(t)\}\leq \bar{H}(t)+\big(1+ 2\beta (n-1)d_t\big)\mathcal{H}(t).
\end{align}

 Hence,  we can invoke Lemma \ref{lem8} to conclude that  $\bar{{H}}(t)$ converges to a nonnegative random variable almost surely as $t$ grows to infinity, which immediately implies that $H(t)$ converges almost surely.

\medskip

\noindent Step 3. The convergence of $h(t)$ follows from a symmetric argument as the analysis to $H(t)$. We have now completed the proof.  \hfill$\blacksquare$

\medskip

\begin{lem}\label{lemr1} Assume that $D_t=0$ for $t=0,\dots, 2(n-2)K-1$. Let $\alpha \in (0,(n-1)^{-1})$ and $i\in\mathcal{V}$.  Then for any $t=0,\dots, 2(n-2)K-1$, there hold

\begin{itemize}
\item[(i)] If $s_i(t)\leq \zeta_0 h(0)+ (1-\zeta_0) H(0)$ for some $\zeta_0\in(0,1)$, then $s_i(t+1)\leq \lambda_\ast \zeta_0  h(0)+(1- \lambda_\ast \zeta_0  )H(0) $, where $\lambda_\ast =1-\alpha(n-1)$;

\item[(ii)] If $s_i(t)\leq \zeta_0 h(0)+ (1-\zeta_0) H(0)$ for some $\zeta_0\in(0,1)$, $B_t=1$, and $(i,j)\in G_t$, then  $s_j(t+1)\leq  \alpha \zeta_0 h(0)+ (1-\alpha \zeta_0) H(0)$.
    \end{itemize}
\end{lem}
{\it Proof.} Note that  the conditions that $D_t=0$ for $t=0,\dots, 2(n-2)K-1$ and  $\alpha \in (0,(n-1)^{-1})$ yield $H(t+1)\leq H(t)$ and $h(t+1)\geq h(t)$ for all $t=0,\dots, 2(n-2)K-1$.

\medskip

\noindent (i). If  $D_t=0$ and $s_i(t)\leq \zeta_0 h(0)+ (1-\zeta_0) H(0)$ for some $\zeta_0\in(0,1)$, then
\begin{align}
s_i(t+1)&=s_i(t)+  \alpha{B}_t h_i^+(t) \nonumber\\
&\leq s_i(t)- \alpha\sum_{j\in{N}_i^+(t) } \big(s_i(t)-s_j(t)\big)\nonumber\\
&\leq (1-\alpha | {N}_i^+(t) |) s_i(t)+\alpha | {N}_i^+(t) | H(t)\nonumber\\
&\leq (1-\alpha | {N}_i^+(t) |)\big( \zeta_0 h(0)+ (1-\zeta_0) H(0)\big) \nonumber\\
&+\alpha | {N}_i^+(t) | H(0)\nonumber\\
& \leq \lambda_\ast \zeta_0  h(0)+(1- \lambda_\ast \zeta_0  )H(0)
\end{align}
in light of the fact that $\alpha \in (0,(n-1)^{-1})$, where $\lambda_\ast =1-\alpha(n-1)$.

\medskip

\noindent (ii) If $s_i(t)\leq \zeta_0 h(0)+ (1-\zeta_0) H(0)$ for some $\zeta_0\in(0,1)$, $B_t=1$, and $(i,j)\in G_t$, there holds that
\begin{align}
s_j(t+1)&=s_j(t)- \alpha\sum_{k\in{N}_j^+(t) } \big(s_j(t)-s_k(t)\big)\nonumber\\
&=(1-\alpha | {N}_j^+(t) |) s_j(t)+\alpha s_i(t) \nonumber\\
& + \alpha\sum_{k\in{N}_j^+(t) \setminus \{i\}} s_k(t)\nonumber\\
&\leq (1-\alpha) H(t)+ \alpha\big(\zeta_0 h(0)+ (1-\zeta_0) H(0)\big)\nonumber\\
&\leq \alpha \zeta_0 h(0)+ (1-\alpha \zeta_0) H(0).
\end{align}
This proves the desired lemma.
\hfill$\blacksquare$

\subsection{Proof of Theorem \ref{thmrsr1}}
\noindent (i). Let $\sum_{t=0}^\infty d_t<\infty$.  Then as long as $\sum_{t=0}^\infty b_t<\infty$, the first Borel-Cantelli Lemma guarantees that almost surely, each node revises its state for only a finite number of slots, which yields  the desired claim follows straightforwardly. In the following, we prove the desired conclusion based on the assumption  that $\sum_{t=0}^\infty b_t=\infty$.

With $\sum_{t=0}^\infty d_t<\infty$, from the first Borel-Cantelli Lemma,
$$
K_\ast:= \inf\{k\geq 0: D_t=0, \forall t\geq k\}
$$
is a finite number almost surely. We note that $K_\ast$ is not a stopping time for $\{D_t\}_{t\geq 0}$, but a stopping time for $\{B_t\}_{t\geq 0}$ by the independence of $\{B_t\}_{t\geq 0}$ and $\{D_t\}_{t\geq 0}$. Hence, we can recursively define
$$
K_{m+1}:= \inf\{t> K_m : B_t=1\},\ m=0,1,\dots
$$
with $K_0:= \inf\{t\ge K_\ast: B_t=1\}$, which are
 are stopping times for $\{B_t\}_{t\geq 0}$. Now in view of the independence of $\{G_t\}_{t\geq 0}$ and $\{B_t\}_{t\geq 0}$, we know  that $\{G_{K_m}\}_{m\geq 0}$ is an independent  process and each $G_{K_m}$ satisfies  $\mathds{P}\big( (i,j)\in E_{K_m}  \big)\geq p_\ast$ for all $(i,j)\in \mathcal{G}$ under Assumption A1.

Let $\mathcal{V}^\dag$ be a positive cluster of $\mathcal{G} $. By assumption, $\mathcal{V}^\dag$ has a spanning tree. Since $\alpha<1/(n-1)$, the above discussion shows that at times $K_m,m=0,1,\dots$, the considered relative-state-flipping model defines a standard  consensus dynamics on independent  random graphs where each arc exists with probability at least $p_\ast$ for any fixed time slot.  Therefore, applying  Theorem 3.4 in \cite{shiIT}  on randomized consensus dynamics with arc-independent graphs, we conclude  that the connectivity of $\mathcal{V}^\dag$ ensures that
$$
\mathds{P}\big(\lim_{m \rightarrow \infty} \mathcal{H}^\dag(K_m)=0\big)=1,
 $$
 where $\mathcal{H}^\dag(t)=\max_{i\in \mathcal{V}^\dag} s_i(t)-\min_{i\in \mathcal{V}^\dag} s_i(t)$. This immediately gives us
 $$
 \mathds{P}\big(\lim_{t \rightarrow \infty} \mathcal{H}^\dag(t)=0\big)=1
  $$
  by the definition of the $K_m$.

  Finally, applying  Lemma \ref{lem10} to the subgraph generated by node set  $\mathcal{V}^\dag$, both $\max_{i\in \mathcal{V}^\dag} s_i(t)$ and $\min_{i\in \mathcal{V}^\dag} s_i(t)$ almost surely converge, and define their limits as, respectively, $H^\dag_\ast$ and $h^\dag_\ast$. Thus, there holds that
  $$
  \mathds{P}\big(\lim_{t \rightarrow \infty} \max_{i\in \mathcal{V}^\dag} s_i(t)= H^\dag_\ast\big)=1
  $$ and that $$
  \mathds{P}\big(\lim_{t \rightarrow \infty} \min_{i\in \mathcal{V}^\dag} s_i(t)= h^\dag_\ast\big)=1.
  $$
  The fact that $\mathds{P}\big(\lim_{t \rightarrow \infty} \mathcal{H}^\dag(t)=0\big)=1$ immediately leads to $H^\dag_\ast=h^\dag_\ast$ almost surely. As a result, we conclude that
  $$  \mathds{P}\big( \lim_{t \rightarrow \infty} s_i(t)= H^\dag_\ast =h^\dag_\ast\big)=1$$  for all $i\in \mathcal{V}^\dag$. This proves the desired statement.

\noindent (ii) Let $ \mathcal{V}_1$ and $\mathcal{V}_2$ be the two positive-clusters  of $\mathcal{G}$. Let $s_i(0)=0, i\in \mathcal{V}_1$ and $s_i(0)=C_0, i\in \mathcal{V}_2$ for some $C_0>0$. We define
$$
f_1(t):= \max_{i\in \mathcal{V}_1} s_i(t);\ \ f_2(t):= \min_{i\in \mathcal{V}_2} s_i(t).
$$
Since the either of the positive cluster contains positive links only and  $\alpha \in (0, (n-1)^{-1})$, there always holds that
$$
f_1(t+1)\leq f_1(t);\ \ f_2(t+1)\geq f_2(t).
$$
Now that there is a negative arc between any two nodes from different clusters, it is straightforward to see that
\begin{align}\label{20}
f_2(t+1)-f_1(t+1)&\geq (1+\beta) \big( f_2(t)-f_1(t) \big)\nonumber\\
& \geq  f_2(t)-f_1(t)+ C_0
\end{align}
whenever $D_t=1$ and either $(i_\ast, j_\ast)\in E_t$ or  $(j_\ast, i_\ast)\in E_t$ with $i_\ast=\arg \max_{i\in \mathcal{V}_1} s_i(t)$ and $j_\ast=\arg \min_{i\in \mathcal{V}_2} s_i(t)$. In light of Assumption A1, the second Borel-Cantelli Lemma (cf., Theorem 2.3.6, \cite{durr}) leads to that the event defined in (\ref{20}) happens infinitely often with probability one when $\sum_{t=0}^\infty d_t=\infty$. The desired conclusion follows immediately.

The proof is now complete. \hfill$\blacksquare$

\subsection{Proof of Theorem \ref{thmrsr2}} The proof relies on  Lemma \ref{lem7},  cf.,  \cite{it10} for the analysis of randomized consensus.

Consider $2n-3$ intervals $[mK,(m+1)K-1],m=0,\dots,2(n-2)$. With Assumption A7, there is a center node $v_m \in \mathcal{V}$ in each of $\mathcal{G}([mK,(m+1)K-1])$. As a result, we can find  $n-1$ center nodes  (repetitions are allowed) out of the $v_m$'s and denote them as $v_{m_1},\dots,v_{m_{n-1}}$, that satisfy  either  or $s_{v_{m_j}}(0) > (h(0)+H(0))/2$, for all $j=1,\dots,n-1$. The two cases are symmetric and without loss of generality, we consider the first case only.

Now we assume that $D_t=0$ for $t=0,\dots, 2(n-2)K-1$. We carry out the following recursive argument:

\begin{itemize}
\item[1)] By our selection  $v_{m_1}$ is a center node of the graph $\mathcal{G}([\tau_1K,(\tau_1+1)K-1])$ for some  $\tau_1=0,\dots,2(n-2)$ with $s_{v_{m_1}}(0) \leq (h(0)+H(0))/2$. Applying Lemma \ref{lemr1}.(i) we conclude that
$$
s_{v_{m_1}}\big(K_0\big)\leq \frac{\rho_\ast^{K_0}}{2} h(0)+ \big(1-\frac{\rho_\ast^{K_0}}{2}\big)H(0),
$$
where $K_0$ and $\rho_\ast$ are defined in the statement of Theorem \ref{thmrsr2}.

\item[2)] Since $v_{m_1}$ is a center, there exist $t_1\in [\tau_1K,(\tau_1+1)K-1]$ and $j_\ast \neq v_{m_1} \in \mathcal{V}$ such that $(v_{m_1}, j_\ast)\in E_{t_1}$ with probability at least $p_\ast$. If $B_{t_1}=1$ and $(v_{m_1}, j_\ast)\in E_{t_1}$, then we can apply Lemma \ref{lemr1} and then conclude
$$
s_{j_\ast}\big(K_0\big)\leq \frac{\rho_\ast^{K_0}}{2} h(0)+ \big(1-\frac{\rho_\ast^{K_0}}{2}\big)H(0).
$$
For convenience we re-denote $v_{m_1}$ and $j_\ast$ as $u_1$ and $u_2$, respectively.

\item[3)] We proceed for $v_{m_2}$. If $v_{m_2}\notin \{u_1,u_2\}$, applying  Lemma \ref{lemr1}.(i) again and we can obtain the same bound for $s_{j_\ast}\big(K_0\big)$. Otherwise either $v_{m_2}=u_1$ or $v_{m_2}=u_2$ allows us to find another node $u_3$ with the bound for $s_{u_3}\big(K_0\big)$ obtained as step 2).
\end{itemize}
From the selection of $v_{m_1},\dots,v_{m_{n-1}}$, the above procedure eventually gives us the same bound for nodes $u_1,\dots,u_N$, and calculating the probability of the required events in the above argument  we obtain
\begin{align}
&\mathds{P}\Big( s_i\big(K_0\big)\leq \frac{\rho_\ast^{K_0}}{2} h(0)+ \big(1-\frac{\rho_\ast^{K_0}}{2}\big)H(0),\ i\in\mathcal{V} \Big) \nonumber\\
&\geq  p_\ast^{n-1} \prod_{t=0}^{K_0-1}\big(b_t(1-d_t)\big). \nonumber
\end{align}
This implies
\begin{align}\label{60}
\mathds{P}\Big( \mathcal{H}\big(K_0\big)\leq  \big(1-\frac{\rho_\ast^{K_0}}{2}\big)\mathcal{H}(0) \Big)\geq  p_\ast^{n-1} \prod_{t=0}^{K_0-1}\big(b_t(1-d_t)\big).
\end{align}

On the other hand, from the definition of the algorithm there always hold
\begin{align}\label{61}
\mathds{P}\big( \mathcal{H}\big(t+1\big)\leq  \big(1+2\beta(n-1) \big)\mathcal{H}(0) \big)= 1
\end{align}
and
\begin{align}\label{62}
\mathds{P}\big( \mathcal{H}\big(K_0\big)>\mathcal{H}(0) \big)\leq 1- \prod_{t=0}^{K_0-1}(1-d_t).
\end{align}

Since $\{{B}_t\}_{t\ge 0}$, $\{{D}_t\}_{t\ge 0}$, and $\{G_t\}_{t\ge 0}$ define independent processes, we conclude from (\ref{60}), (\ref{61}), and (\ref{62}) that
\begin{align}
\mathds{E}\big\{\mathcal{H}\big((m+1)K_0\big)\big|\mathcal{H}\big(m K_0\big)  \big\} \leq \big(1- X_m+ Y_m\big)\mathcal{H}\big(m K_0\big). \nonumber
\end{align}
The desired theorem then follows directly from Lemma \ref{lem7} and (\ref{61}). \hfill$\blacksquare$

\subsection{Proof of Theorem \ref{thmrsr3}}
In light of  $\alpha \in [0,(n-1)^{-1}/2)$, we first prove two claims.

\medskip

\noindent {\it Claim A.} $\mathds{P}\Big(\mathcal{H}(t+1)\geq \big(1-2(n-1)\alpha\big)\mathcal{H}(t)\Big)=1$.

\medskip

\noindent {\it Claim B.} $\mathds{P}\big(\mathcal{H}(t+1)<\mathcal{H}(t)\big)\leq b$.

Take $i,j\in\mathcal{V}$ satisfying $s_i(t)=h(t)$ and $s_j(t)=H(t)$. Similarly as the proof of Lemma \ref{lemr1}, we can establish that almost surely,
\begin{align}\label{51}
s_i(t+1)\leq  \lambda_\ast h(t)+ (1-\lambda_\ast) H(t)
\end{align}
and
\begin{align}\label{52}
s_j(t+1)\geq  \lambda_\ast H(t)+ (1-\lambda_\ast) h(t)
\end{align}
hold, where $ \lambda_\ast=1-\alpha(n-1)$. Noting that (\ref{51}) and (\ref{52}) yield
\begin{align}
\mathcal{H}(t+1)&\geq |s_j(t+1)-s_i(t+1)|\nonumber\\
&\geq |2\lambda_\ast-1|\mathcal{H}(t)\nonumber\\
&=\big(1-2(n-1)\alpha\big)\mathcal{H}(t),
\end{align}
Claim A is proved.

Furthermore, if $b_t=0$, only negative recommendations can be effective in the node state update. This implies Claim B.

Now we define $L_0:=\inf \{t\in \mathds{Z}: (1+\beta)^t\geq 2(n-1)\}$. Consider time intervals $[mK,(m+1)K-1]$ for $m=0,1,\dots, (n^2-n)(L_0-1)$. Denote $K_{L_0} =K((n^2-n)(L_0-1)+1)$. Under Assumption A8 and based on the fact that there are at most $n(n-1)$ arcs, there are two nodes $i_\ast,j_\ast \in \mathcal{V}$ and $L_0$ instants $0\leq \tau_1<\tau_2<\dots<\tau_{L_0}<K_{L_0}$    such that $(i_\ast,j_
\ast) \in\mathcal{G}^-_{\tau_k}$ and $|s_{i_\ast}(\tau_k)-s_{j_\ast}(\tau_k)|\geq \mathcal{H}(\tau_k)/(n-1)$ for all $\tau_k$.

Consider the following event:
\begin{align}
&\mathrm{E}_\ast:=\Big\{ D_{\tau_k}=1, i_\ast= N^-_{j_\ast} (\tau_k) {\rm\ for\ all\ } \tau_k; \nonumber\\
&  B_t=0 {\rm\ for\ all\ } t\in[0,K_{L_0}-1]\Big\}.
\end{align}
The event $\mathrm{E}_\ast$ implies
$$
\mathcal{H}(K_{L_0})\geq |s_{i_\ast}(K_{L_0})-s_{j_\ast}(K_{L_0})|\geq \mathcal{H}(0)(1+\beta)^{L_0} \cdot (n-1)^{-1}.
$$
As a result, we can bound the probability of $\mathrm{E}_\ast$ and conclude
\begin{align}\label{54}
&\mathds{P}\Big(\mathcal{H}(K_{L_0})\geq \mathcal{H}(0)(1+\beta)^{L_0} \cdot (n-1)^{-1}\Big)\nonumber\\
&\geq \big( d p_\ast(1-p^\ast)^{n-2}\big)^{L_0} (1-b  )^{K_{L_0}}.
\end{align}

We can now apply the same argument as  the proof of Proposition 1 in \cite{P1}. With (\ref{54}), Claim A, and Claim B, there holds
\begin{align}
&\mathds{E}\Big\{  \log \mathcal{H}(K_{L_0}) - \log\mathcal{H}(0)\Big\} \geq \big( d p_\ast(1-p^\ast)^{n-2}\big)^{L_0} (1-b  )^{K_{L_0}}\nonumber\\
 &\cdot \log\Big((1+\beta)^{L_0} \cdot (n-1)^{-1}\Big)+b\log \big(1-2(n-1)\alpha\big)\nonumber\\
&\geq \big( d p_\ast(1-p^\ast)^{n-2}\big)^{L_0} (1-b  )^{K_{L_0}} \log 2\nonumber\\
&+b\log \big(1-2(n-1)\alpha\big)\nonumber\\
&>0
\end{align}
when $b<b_\star$ for some sufficiently small $b_\star>0$. We can proceed to define $U(m)=\log \mathcal{H}(mK_{L_0})$ for $m=0,1,\dots$. Recursively applying the above arguments to the process $\{U_m\}$ we obtain that $U(m)$ has a strictly positive drift when $b<b_\star$, which  implies that $\liminf_{m\to \infty} U(m)=\infty$ holds almost surely.

This completes the proof. \hfill$\blacksquare$

\subsection{Proof of Theorem \ref{thmrsr4}} Let us focus on a given positive cluster  $\mathcal{V}^\dag$ of $\mathcal{G}$. We use the following notations
$$
\Psi(t)=\max_{i\in \mathcal{V}^\dag} s_i(t), \psi(t)=\min_{i\in \mathcal{V}^\dag} s_i(t), \Theta(t)= \Psi(t)-\psi(t).
$$
Applying  Lemma \ref{lem10} on the positive cluster  $\mathcal{V}^\dag$, we conclude that each of  $\Theta(t)$, $\Psi(t)$, and $\psi(t)$ converge to a finite limit almost surely if $\sum_{t=0}^\infty d_t<\infty$.

In light of Assumption A9, applying the same argument we used in order to establish (\ref{60}) of Theorem \ref{thmrsr2} on the cluster $\mathcal{V}^\dag$, we similarly have
\begin{align}\label{56}
&\mathds{P}\Big( \Theta\big((m+1)K_0\big)\leq  \big(1-\frac{\rho_\ast^{K_0}}{2}\big)\Theta(mK_0) \Big)\nonumber\\
&\geq  p_\ast^{n-1} \prod_{t=mK_0}^{(m+1)K_0-1}\big(b_t(1-d_t)\big).
\end{align}
Moreover, from the effect of the  negative recommendations on nodes in $\mathcal{V}^\dag$, we can easily modify (\ref{61}) and (\ref{62}) to that for all $t$,
\begin{align}\label{58}
\mathds{P}\big( \Theta\big(t+1\big)>\Theta(t) \big)\leq d_t
\end{align}
and
\begin{align}\label{57}
\mathds{P}\Big( \Theta\big(t+1\big)\leq  (1+2\beta (n-1))\mathcal{H}(t)\Big)=1.
\end{align}

With (\ref{56}), (\ref{58}) and (\ref{57}), we arrive at
\begin{align}\label{99}
&\mathds{E}\big\{\Theta\big((m+1)K_0\big)\big|\Theta \big(m(K_0)\big)  \big\} \nonumber\\
&\leq  \big(1-X_m\big)
\Theta\big(mK_0\big) \nonumber\\
&+(1+2\beta(n-1))\sum_{t=mK_0}^{(m+1)K_0-1} d_t \mathcal{H}(t),
\end{align}
where $X_m$ is defined in Theorem \ref{thmrsr2}.

On the other hand, from (\ref{100}) we know that $$
\mathds{E}(\mathcal{H}(t))\leq  \mathcal{H}_0\prod_{t=0}^\infty\big(1+ 2\beta (n-1)d_t\big)
$$ for all $t\geq 0$. Taking the  expectation from the both sides of  (\ref{99}), we obtain:
\begin{align}
&\mathds{E}\big\{\Theta\big((m+1)K_0\big)\big\} \leq  \big(1-X_m\big)
\mathds{E}\big\{ \Theta\big(mK_0\big)\big\}\nonumber\\
&  + \Big[(1+2\beta(n-1))\mathcal{H}_0\prod_{t=0}^\infty\big(1+ 2\beta (n-1)d_t\big)\Big]W(m),
\end{align}
where $W(m)=\sum_{t=mK_0}^{(m+1)K_0-1} d_t$.

Note that it is well known that  $\sum_{t=0}^\infty d_t   <\infty$ implies $\prod_{t=0}^{\infty}(1-d_t)>0$ and $\prod_{t=0}^\infty\big(1+ 2\beta (n-1)d_t\big)<\infty$. Consequently,  $\sum_{m=0}^\infty J(m)=\infty$ implies $\sum_{m=0}^\infty X_m=\infty$. In view of Lemma \ref{lem7}, we have
\begin{align}
\lim_{m\rightarrow \infty}\mathds{E}\big\{\Theta\big((m+1)K_0\big)\big\}=0
 \end{align}
  if $\lim_{m \rightarrow \infty} W(m)/J(m)=0$. Invoking Fatou's lemma (e.g., Theorem 1.6.5, \cite{durr}), we further conclude that
\begin{align}
\mathds{E}\big\{\liminf_{m\rightarrow \infty} \Theta\big((m+1)K_0\big)\big\}\leq \lim_{t\rightarrow \infty}\mathds{E}\big\{\Theta\big((m+1)K_0\big)\big\}=0,
  \end{align}
which actually implies
\begin{align}
\mathds{E}\big\{\lim_{m\rightarrow \infty} \Theta\big((m+1)K_0\big)\big\}=0
 \end{align}
 since $\Theta\big((m+1)K_0\big)$ converges almost surely. Therefore, we have reached
\begin{align}
\mathds{P}\big(\lim_{m\rightarrow \infty} \Theta\big((m+1)K_0\big)=0\big)=1,
 \end{align}
 which in turn leads to
\begin{align}
\mathds{P}\big(\lim_{t\rightarrow \infty} \Theta(t)=0\big)=1,
 \end{align}
 again, from the fact that $\Theta(t)$ converges almost surely.

Finally,  $\Theta(t)$ converging  almost surely to zero  means that $\Psi(t)$ and $\psi(t)$ must converge to the same limit (their convergence is established in the beginning of the proof), which must be the limit of the each node state in $\mathcal{V}^\dag$. We have now  completed the proof.
  \hfill$\blacksquare$

\section{Conclusions}
This paper continued the study of \cite{shiJSAC,shiOR} investigating a relative-state-flipping model for consensus dynamics over signed random networks.  A sequence of deterministic signed graphs define potential node  interactions that happen independently but not necessarily i.i.d.   The positive recommendations are consistent with the standard consensus algorithm;   negative recommendations are defined by relative-state flipping  from its negative neighbors. Each node puts a (deterministic) weight to each recommendation, and then encodes these weighted recommendations in its state update through stochastic attentions defined by two Bernoulli random variables. We  have established several fundamental  conditions regarding almost sure convergence and  divergence  of the network states.  A condition for  almost sure  state clustering was also proposed for weakly balanced graphs, with the help of  martingale convergence lemmas. Some fundamental differences were also highlighted between the current relative-state-flipping model and the state-flipping model considered in \cite{P1,altafini13}.

\end{document}